\documentclass[%
 twocolumn,
superscriptaddress,
nofootinbib,
 amsmath,amssymb,
 aps,
 prl
]{revtex4-2}

\usepackage{graphicx}
\usepackage{dcolumn}
\usepackage{bm}
\usepackage{siunitx}
\usepackage{textgreek}
\usepackage{newfloat}
\usepackage{mathtools}
\usepackage{nicefrac}
\usepackage{comment}
\usepackage[svgnames]{xcolor}
\usepackage{hyperref}
\hypersetup{colorlinks=true,allcolors=DarkBlue}

\DeclarePairedDelimiter\ket{\lvert}{\rangle}
\DeclareSIUnit\angstrom{\text {Å}}

\begin{document}

\title{Magnetic Ni-N-Ni Centers in N-substituted NiO}

\author{Simon Godin\textsuperscript{*}}
\affiliation{Department of Physics and Astronomy, University of British Columbia, Vancouver, British Columbia V6T 1Z4, Canada}
\affiliation{Quantum Matter Institute, University of British Columbia, Vancouver, British Columbia V6T 1Z4, Canada}

\author{Ilya S. Elfimov}
\affiliation{Department of Physics and Astronomy, University of British Columbia, Vancouver, British Columbia V6T 1Z4, Canada}
\affiliation{Quantum Matter Institute, University of British Columbia, Vancouver, British Columbia V6T 1Z4, Canada}

\author{Fengmiao Li}
\affiliation{Department of Physics and Astronomy, University of British Columbia, Vancouver, British Columbia V6T 1Z4, Canada}
\affiliation{Quantum Matter Institute, University of British Columbia, Vancouver, British Columbia V6T 1Z4, Canada}

\author{Bruce A. Davidson}
\affiliation{Department of Physics and Astronomy, University of British Columbia, Vancouver, British Columbia V6T 1Z4, Canada}
\affiliation{Quantum Matter Institute, University of British Columbia, Vancouver, British Columbia V6T 1Z4, Canada}

\author{Ronny Sutarto}
\affiliation{Canadian Light Source, Saskatoon, Saskatchewan S7N 2V3, Canada}

\author{Jonathan D. Denlinger}
\affiliation{Advanced Light Source, Lawrence Berkeley National Laboratory, Berkeley, California 94720, USA}

 \author{Liu Hao Tjeng}
\affiliation{Max Planck Institute for Chemical Physics of Solids, Nöthnitzer Straße 40, 01187 Dresden, Germany }

\author{George A. Sawatzky}
\affiliation{Department of Physics and Astronomy, University of British Columbia, Vancouver, British Columbia V6T 1Z4, Canada}
\affiliation{Quantum Matter Institute, University of British Columbia, Vancouver, British Columbia V6T 1Z4, Canada}

\author{Ke Zou}
\affiliation{Department of Physics and Astronomy, University of British Columbia, Vancouver, British Columbia V6T 1Z4, Canada}
\affiliation{Quantum Matter Institute, University of British Columbia, Vancouver, British Columbia V6T 1Z4, Canada}

\date{\today}

\begin{abstract}
	
\normalsize

To explore how anion substitution modifies the existing magnetism in strongly correlated oxides, we investigate local electronic states and magnetic ordering in nickel oxide (NiO) induced by substituting oxygen (O) with nitrogen (N). Each N introduces an additional N 2p hole and modifies the magnetic moment of a neighboring nickel (Ni) cation site, as the exchange interaction between this hole and the Ni e\textsubscript{g} electrons exceeds the Ni-O-Ni superexchange interaction. This leads to the formation of Ni-N-Ni centers consisting of five spins, without perturbing the antiferromagnetic NiO lattice. These centers are studied using density functional theory and confirmed through high-resolution spectroscopy on N-substituted NiO thin films grown by molecular beam epitaxy. This type of magnetic design may enable future advances in quantum technologies based on strongly correlated materials, such as quantum sensors and spin-based qubits. \\

\noindent \small DOI: \href{https://doi.org/10.1103/mp1d-jg6c}{10.1103/mp1d-jg6c}

\end{abstract}

\maketitle

\thispagestyle{plain}
\pagestyle{plain}
\footnotetext{Contact author: sgodin@phas.ubc.ca}

The doping of transition metal oxides has been a broad area of study with significant fundamental interest and numerous practical applications. Nickel monoxide (NiO) in particular is an intriguing material, often regarded as a prototypical strongly correlated system \cite{Kunes2007b}. NiO is a charge-transfer insulator with a 4 eV bandgap \cite{Chatterji2009}, located in the intermediate regime of the Zaanen-Sawatzky-Allen scheme \cite{Sawatzky1985}, exhibiting strong hybridization between Ni 3d and O 2p. It has an antiferromagnetic (AFM) rock-salt structure with a lattice parameter a = \SI{4.176}{\angstrom} \cite{Alders1998a} and a Néel temperature (T\textsubscript{N}) of 523 K \cite{Arai2012b}. The AFM superexchange coupling of the Ni next nearest neighbors leads to ferromagnetic (FM) (111) planes with multiple domain walls and spin domains at the surface \cite{Weber2003,Arai2012b}. The replacement of Ni cations with other 3d transition metals, such as manganese (Mn) and cobalt (Co), which varies the number of 3d electrons, has been shown to significantly impact magnetic properties \cite{Park2020,Thi2015}.

\begin{figure*}[t]
	\includegraphics[width = 0.9\textwidth]{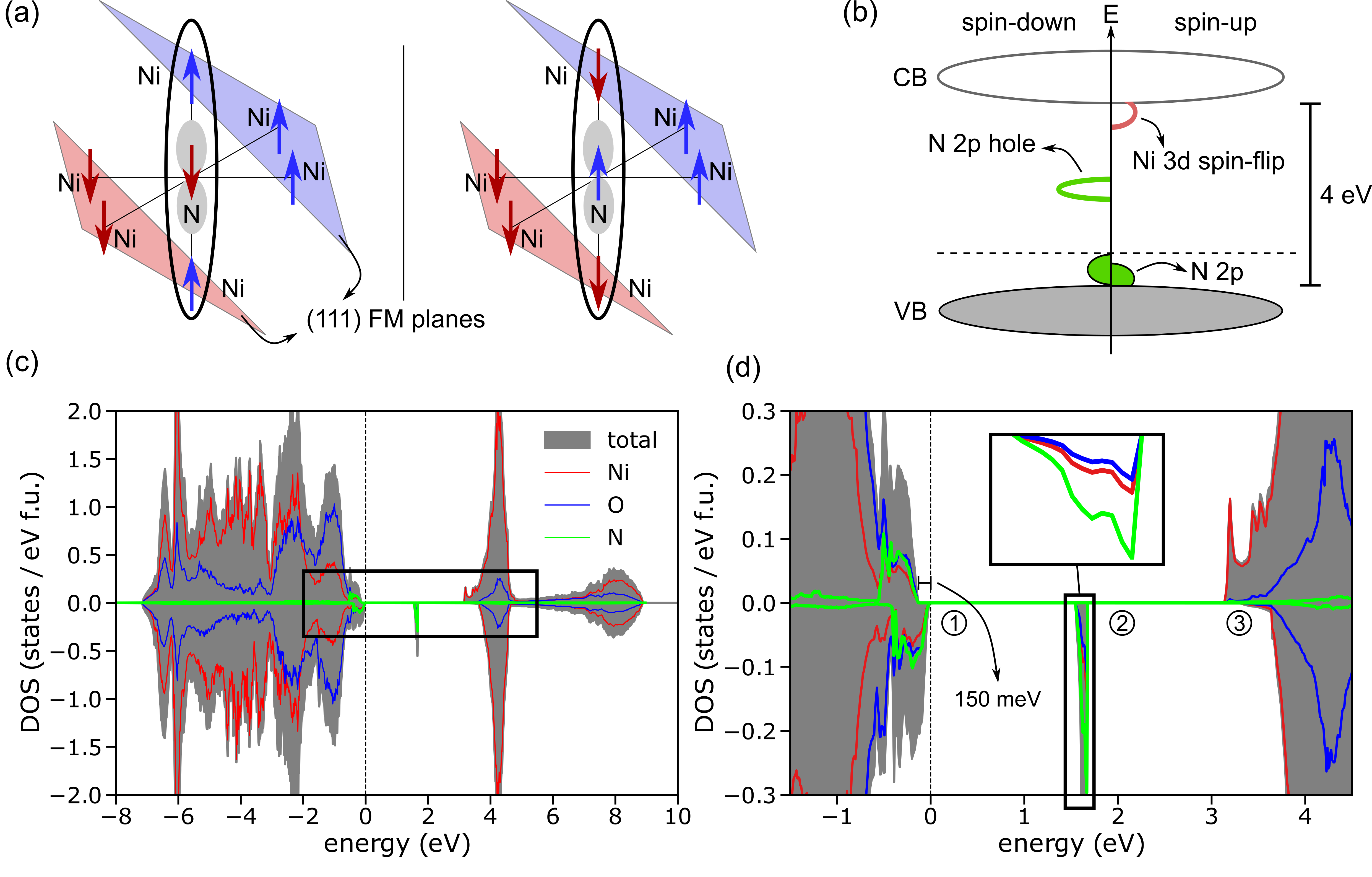}
	\caption{\label{fig:groundstate} Density functional theory (DFT) calculations of N-substituted NiO, with one out of thirty-two O atoms replaced by an N. (a) Representation of the Ni-N-Ni center. The presence of the N flips the spin of a neighboring Ni within its (111) FM plane. The states with an N 2p hole aligned spin-down (or spin-up) and both Ni 3d aligned spin-up (or spin-down) are degenerate. (b) Schematic of the introduced PDOS near the bandgap of NiO upon N substitution. (c) PDOS of Ni, O, and N resulting from this configuration. (d) Close-up of the energy window around the bandgap of NiO\textsubscript{1-x}N\textsubscript{x} showing N, Ni, and O's PDOS. The labels \textbf{(1)}, \textbf{(2)}, and \textbf{(3)} point to the shoulder to the valence band, impurity states, and shoulder to the conduction band, respectively. These features are shown in (b). The zero of energy represents the Fermi energy.
	}
\end{figure*}

What remains intriguing, yet insufficiently explored, is the effect of anion substitution in NiO, such as nitrogen (N). It is expected to significantly alter local electronic and magnetic interactions, considering the more extensive 2p orbital and the one less electron in N compared to O. Previous investigations of N-substituted nonmagnetic oxides, such as SrO\textsubscript{1-x}N\textsubscript{x} \cite{Elfimov2007a} and ZnO\textsubscript{1-x}N\textsubscript{x} \cite{Shen2008,Jindal2012}, report that N incorporation induces FM order, with magnetic moments localized primarily on the anion sites. In contrast, N-substituted NiO presents a fundamentally different scenario: it not only hosts intrinsic AFM order, but also exhibits N-induced modifications of magnetic moments on the cation (Ni) sites. This unique interplay between anion substitution and cation spin reconfiguration sets NiO apart, revealing a new mechanism for tailoring magnetic interactions in strongly correlated oxides.

Anion substitution using N and carbon (C) in various materials is used in photonics and quantum information. Key among them are quantum sensors, which enable the detection of single electron and nuclear spins in nanoscale-resolution scanning microscopy and also serve as spin qubits \cite{Taylor2008, Ho2021, Weber2010}. The most common quantum sensors are nitrogen-vacancy (NV) centers, consisting of a neighboring C vacancy and an N-substituted C site in diamond. Efforts are ongoing to extend spin centers to other materials, including T centers in silicon (Si) \cite{Higginbottom2022, Bergeron2020}, and defects in hexagonal boron nitride (hBN) \cite{Stern2024} and SiC \cite{Wolfowicz2020}.

In this Letter, we demonstrate experimentally that incorporating N in NiO thin films induces in-gap states of N 2p character. Since this hole remains localized at the anion site, N adopts an N\textsuperscript{2-} oxidation state instead of N\textsuperscript{3-}. We calculate that the substitution of an O with N flips the spin of a neighboring Ni, forming a Ni-N-Ni center with total spin $S = 3/2$ in the ground state. This configuration emerges due to a strong AFM exchange interaction between the N 2p hole and the Ni electrons, which in turn leads to FM spin alignment between the two Ni sites. This distinct phenomenon establishes a new paradigm for locally tuning magnetic order through anion substitution in strongly correlated oxides and gives rise to a local spin that behaves like a paramagnetic, half-integer spin impurity.

We validate these results through both density functional theory (DFT) calculations and spectroscopy experiments on NiO\textsubscript{1-x}N\textsubscript{x} thin films grown by molecular beam epitaxy (MBE). Using experimental and calculated X-ray absorption spectroscopy (XAS), we provide direct evidence of a change in the magnetic property of Ni and the presence of N impurity states in the bandgap. We also observe the emergence of states below the Fermi energy using angle-resolved photoemission spectroscopy (ARPES) and confirm the N character of the introduced holes through X-ray photoelectron spectroscopy (XPS). Anion-substituted NiO prepared using MBE is stable in air, and can be handled \textit{ex situ}, making it an attractive candidate for practical applications.

Figure \ref{fig:groundstate} shows DFT+U calculations of N-substituted NiO (see calculation methods in Supplemental Material \cite{SM_PRL}). The electron's on-site effective interactions for the Ni 3d, O 2p, and N 2p states are U = 6, 4, and 3 eV, respectively. Our calculations indicate that the substitution of O atoms with N introduces one hole of N 2p character, allowing the oxidation state of the Ni atoms to remain Ni\textsuperscript{2+}. The arrangement in which each N atom flips the spin of a neighboring Ni within its FM plane [see Fig. \ref{fig:groundstate}(a)] significantly lowers the energy of the system by 204 meV. Notably, the rest of the AFM order and overall structure are preserved throughout the rest of the material.

\nocite{Lin2013a,Sriram2016b,Wicks2012b,Voogt2001a,Godin2022,Bjorck2007,QE-2009,QE-2017,PAW-1994,PBE-1996,DFTU-1991,DFTU-1998,kedges1,kedges2,Achkar2011a,Herrera-Gomez2014} 

Figure \ref{fig:groundstate}(b) schematically shows the main features of the resulting spin-resolved partial density of states (PDOS) that depart from unsubstituted NiO. The N 2p hole creates spin-polarized impurity states in the bandgap, while the spin-flipped Ni forms a shoulder below the conduction band with opposite spin. Additionally, spin-polarized states appear above the valence band.

Figure \ref{fig:groundstate}(c) shows the computed PDOS results of N-substituted NiO in the spin-flipped Ni configuration, and Fig. \ref{fig:groundstate}(d) highlights the bandgap. At the top of the valence band \textbf{(1)}, spin-down states shift by approximately 150 meV relative to spin-up due to strong kinetic exchange between N 2p and neighboring Ni 3d orbitals. Notably, this is almost an order of magnitude stronger than the $\sim 20$ meV superexchange interaction between Ni 3d through the full O 2p orbitals in NiO \cite{DeGraaf1997,Chatterji2009,Hutchings_1972}. We also observe the impurity states \textbf{(2)} dominated by N 2p and the shoulder to the conduction band \textbf{(3)}. We show in Supplemental Material \cite{SM_PRL} that flipping the spin of a Ni in unsubstituted NiO costs instead 117 meV, and present the energetically unfavorable ground state of NiO\textsubscript{1-x}N\textsubscript{x} with no spin-flipped Ni.

In unsubstituted NiO, the AFM superexchange interaction between Ni atoms, mediated by an O atom, results in two triangular arrangements of Ni spins in an AFM configuration. The large exchange interaction between the Ni 3d e\textsubscript{g} electrons leads to a net spin moment $S=1$ at each Ni site, as dictated by Hund's rules. Replacing one O by N introduces an extra spin $S = 1/2$ at the anion site due to the absence of a 2p electron. The N spin has an AFM exchange interaction with either of the nearest neighbor Ni atoms it points to, which replaces the Ni-O-Ni superexchange.

For instance, a 2p\textsubscript{z}-oriented N orbital will couple exclusively with the 3d\textsubscript{3z2-r2} orbitals of the Ni cations. We can estimate the hopping integral $t_{pd} \approx 1.5$ eV between 2p\textsubscript{z} and 3d\textsubscript{3z2-r2}. Considering the charge-transfer energy $\Delta \approx 4$ eV [see Fig. \ref{fig:groundstate}(b)] and the Coulomb interaction between the two N holes $U_{pp} \approx 3$ eV, we find:
\begin{equation}
J_{pd} = \frac{t_{pd}^2}{\Delta + U_{pp}} \approx 300\, \text{meV},
\end{equation}
with a \SI{180}{\degree} bond between the N and each Ni.

This Ni-N AFM interaction is larger than the original superexchange interaction of $\sim 20$ meV and, in turn, leads to an FM spin orientation between the two Ni via double exchange. The Hund's rule coupling $J_H \approx 1$ eV \cite{Terakura1984}, which gives rise to the spin-1 state of Ni, is significantly larger than $J_{pd}$, resulting in a ground state with total spin $S_{tot} = 3/2$. The value of $J_{pd}$ we find here is larger than the one resulting from the DFT calculations ($ J_{pd} \approx $ 150 meV), due to DFT underestimating the exchange interaction between specific orbitals.

Within this theoretical framework, the Ni-N-Ni center is found to be magnetically decoupled from the surrounding AFM lattice at zero temperature. This results in degenerate configurations where two Ni ions are spin-aligned (up/down) and the N atom carries an opposing spin (down/up). This degeneracy can be lifted by applying an external magnetic field. This suggests that the Ising spin configuration assumed in DFT is insufficient, as the NiO zero-point spin fluctuations prevent the local moments from maintaining well-defined $S_z$ eigenstates, instead reducing them. Nevertheless, further investigation is needed to assess the robustness of this decoupling. Perturbations such as lattice distortion could reintroduce coupling to the host lattice, potentially lifting the degeneracy even at zero temperature and without an external magnetic field.

We study these effects experimentally in NiO\textsubscript{1-x}N\textsubscript{x} thin films grown by MBE using NO gas on MgO (001) substrates. The N-substituted NiO films retain the same rock-salt crystal structure as NiO control samples. We show in Supplemental Material \cite{SM_PRL} their high crystallinity, confirmed using reflection high-energy electron diffraction and high-resolution X-ray diffraction, and discuss the growth details.

\begin{figure}[t]
	\includegraphics[width = \columnwidth]{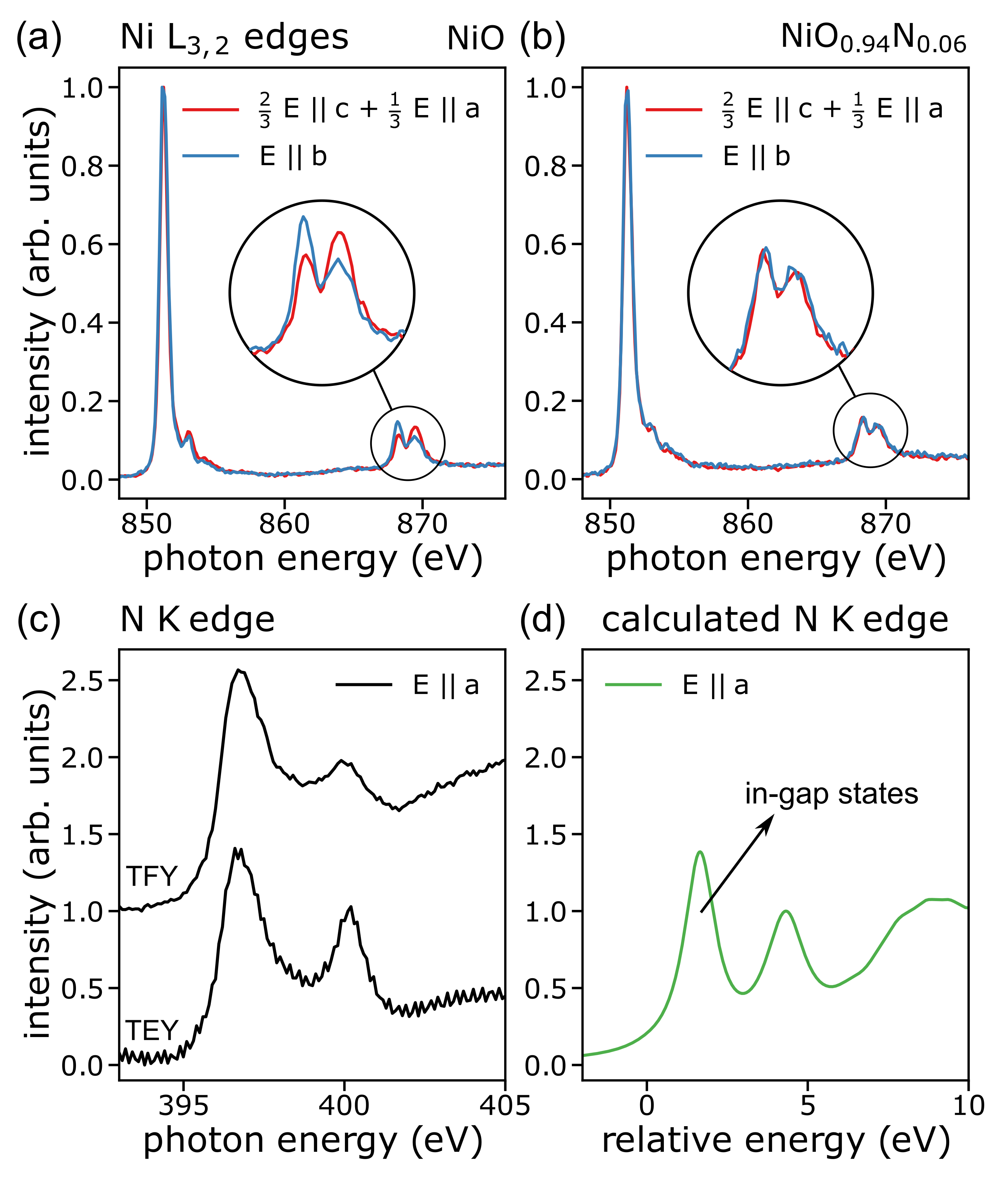}
	\caption{\label{fig:XAS} X-ray absorption spectroscopy (XAS) of NiO\textsubscript{1-x}N\textsubscript{x}. Ni L\textsubscript{3,2}-edges of (a) NiO and (b) NiO\textsubscript{0.94}N\textsubscript{0.06}, at grazing incidence. The change in dichroism implies that the magnetic order is affected upon substitution. The spectra are taken in IPFY mode at the O 1s emission energy. (b) Experimental N K-edge of NiO\textsubscript{0.94}N\textsubscript{0.06} in TEY and TFY modes, at normal incidence. The first peak arises from the in-gap states. (d) Calculated N K-edge from the DFT results. A 1 eV Lorentzian broadening is applied to replicate the experimental results.
	}
\end{figure}

\begin{figure}[t]
	\includegraphics[width = \columnwidth]{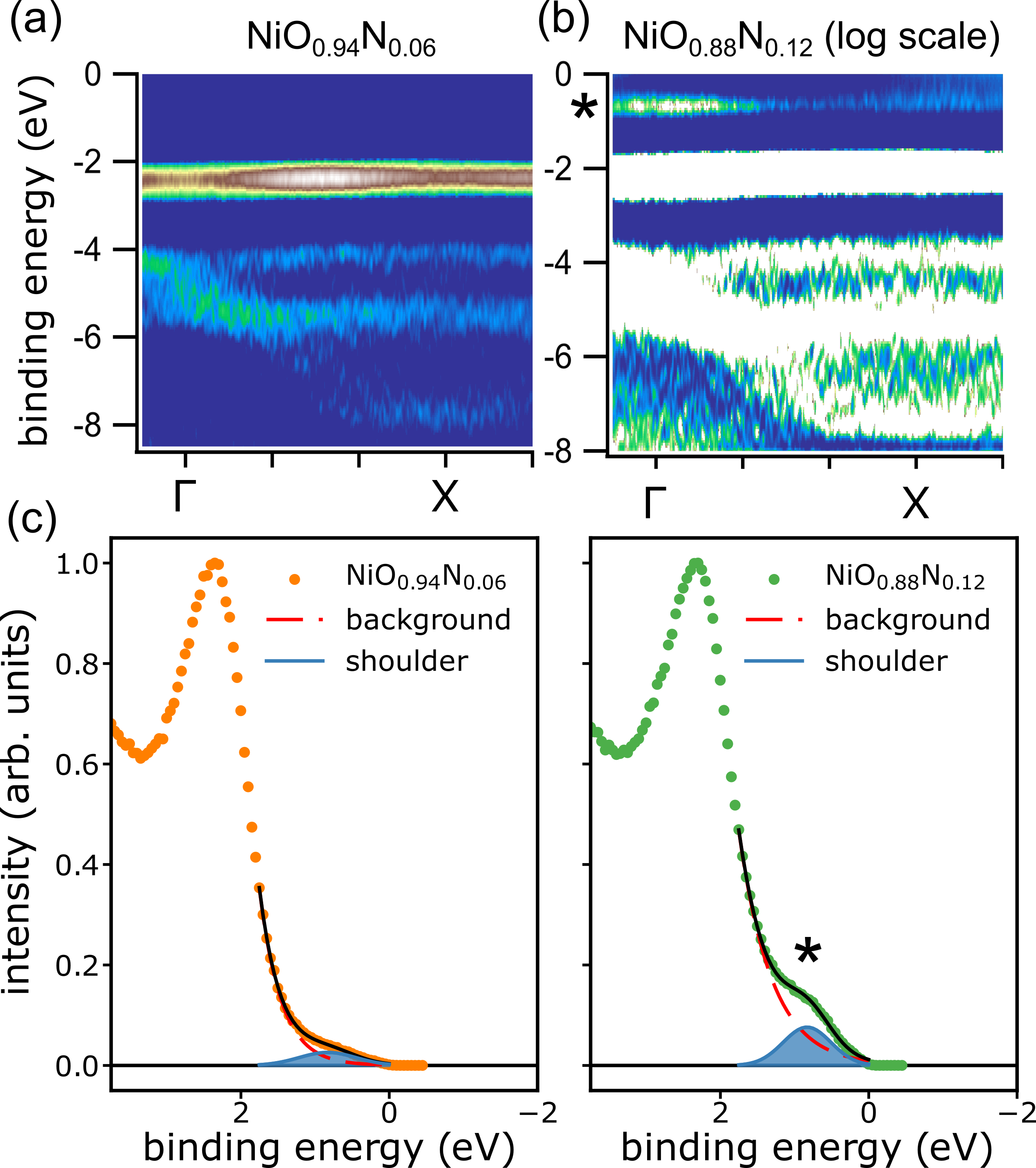}
	\caption{\label{fig:ARPES} Angle-resolved photoemission spectroscopy (ARPES) of NiO\textsubscript{1-x}N\textsubscript{x} with incoming photon energy of 85 eV. (a) Momentum distribution curve (MDC) 2\textsuperscript{nd} derivative of $\Gamma$-$X$ cuts in NiO\textsubscript{0.94}N\textsubscript{0.06}. (b) MDC 2\textsuperscript{nd} derivative in logarithmic scale of $\Gamma$-$X$ cuts in NiO\textsubscript{0.88}N\textsubscript{0.12}. The N 2p shoulder emerges around $\Gamma$ and $X$, and shows minimal dispersion. (c) Comparison of the normalized valence band of 6 and 12\% N-substituted NiO. The shoulder at the top of the valence band arises from the N 2p states and shows a 1:2 intensity ratio between 6 and 12\%.
	}
\end{figure}

Figure \ref{fig:XAS} presents XAS measurements on NiO and NiO\textsubscript{0.94}N\textsubscript{0.06} taken at room temperature. The Ni L\textsubscript{3,2}-edges are resolved using inverse partial fluorescence yield (IPFY) [see Fig. \ref{fig:XAS}(a)], with linear dichroism at the L\textsubscript{2}-edge arising from the long-range AFM order and local crystal field effects in NiO \cite{Alders1998a,Krishnakumar2007a,DeGroot1995a}. The impact of the AFM order on the dichroism intensity is dominant in thick epitaxial films, as those studied here \cite{Haverkort2004}. The spin orientation is influenced by strain, with the $\sim 0.85\%$ tensile strain in NiO on MgO aligning spins out-of-plane relative to the (001) surface \cite{Altieri2003a}. The dichroism in Fig. \ref{fig:XAS}(a), measured at grazing angle $\theta$ = \SI{30}{\degree}, confirms the out-of-plane spin component in NiO. Figure \ref{fig:XAS}(b) shows that this dichroism is suppressed completely upon N substitution. This supports the Ni spin-flip effect [see Fig. \ref{fig:groundstate}(a)], which would hinder the dichroism. Complete, rather than partial, suppression can be attributed to each N having 12 nearest neighbors in the FCC anion lattice. This results in 50\% of the N having a next nearest neighbor N \cite{Behringer1958}, potentially enhancing magnetic disorder.

Figure \ref{fig:XAS}(c) shows the N K-edge spectra of a 6\% substituted sample in total electron yield (TEY) and total fluorescence yield (TFY) modes at normal incidence. A prominent double peak, also found in nickel nitrides \cite{Pandey2021,You2017,Jin2019}, is present at 396.7 and 400.1 eV. The DFT-calculated absorption spectrum of Fig. \ref{fig:XAS}(d) under a 1 eV Lorentzian broadening shows good agreement between the experimental and DFT results. Supplemental Material \cite{SM_PRL} includes the 2p\textsubscript{z}, 2p\textsubscript{y}, 2p\textsubscript{x} contributions without broadening. The impurity states seen $\sim 1.8$ eV above the Fermi energy in Fig. \ref{fig:groundstate}(d) form the first peak, while the second corresponds to hybridized N 2p and Ni 3d states. The in-gap states are also probed at the O K-edge, forming a shoulder to the pre-edge [see Supplemental Material \cite{SM_PRL}]. The comparable N K-edge spectra observed in both TEY and TFY modes indicate a similar behavior between the surface layers and the bulk.

We use \textit{ex situ} ARPES measurements to reveal the band structure of N-substituted NiO. In NiO, such studies are limited by charging effects \cite{Gillmeister2020,Shen1991}, but the lower resistivity of the N-substituted samples mitigates this issue. The ground state of NiO, with hybridized Ni 3d and O 2p orbitals, can be expressed as $\alpha \ket{3\text{d}^8} + \beta \ket{3\text{d}^9\underline{\text{L}}} + \gamma \ket{3\text{d}^{10}\underline{\text{L}}^2}$ \cite{VanVeenendaal1993}, where \underline{L} denotes a ligand hole. This results in a complex spectrum spawning from 0 to -8 eV. We show in Supplemental Material \cite{SM_PRL} our ability to observe clear constant-energy maps, further confirming the high crystallinity and resilience to air exposure of the substituted samples. The scans are taken at 85 eV and grazing incidence ($\theta$ = \SI{25}{\degree}). Figure \ref{fig:ARPES}(a) shows the cuts along $\Gamma$ to $X$ of NiO\textsubscript{0.94}N\textsubscript{0.06}. The high-intensity Ni 3d band is seen at -2.2 eV, while the hybridized states of O 2p and Ni 3d characters spread down to -8 eV.

The N 2p states create a new band around -0.75 eV, in agreement with the DFT calculations in Fig. \ref{fig:groundstate}(d). We resolve this band in Fig. \ref{fig:ARPES}(b) by looking at the 2\textsuperscript{nd} derivative of the $\Gamma$-$X$ cut with a logarithmic scale in intensity in a 12\% substituted sample, which shows limited dispersion and a larger weight around $\Gamma$ and $X$. This feature is seen clearly in Fig. \ref{fig:ARPES}(c) as a shoulder in the angle-integrated valence band. The intensity ratio of these states from 6 to 12\% substitution is 2$\pm$0.4, with both showing a peak separation of $\sim 1.5$ eV between the energy at the highest intensity of the valence band and its shoulder.

\begin{figure}[b!]
	\includegraphics[width = 0.75\columnwidth]{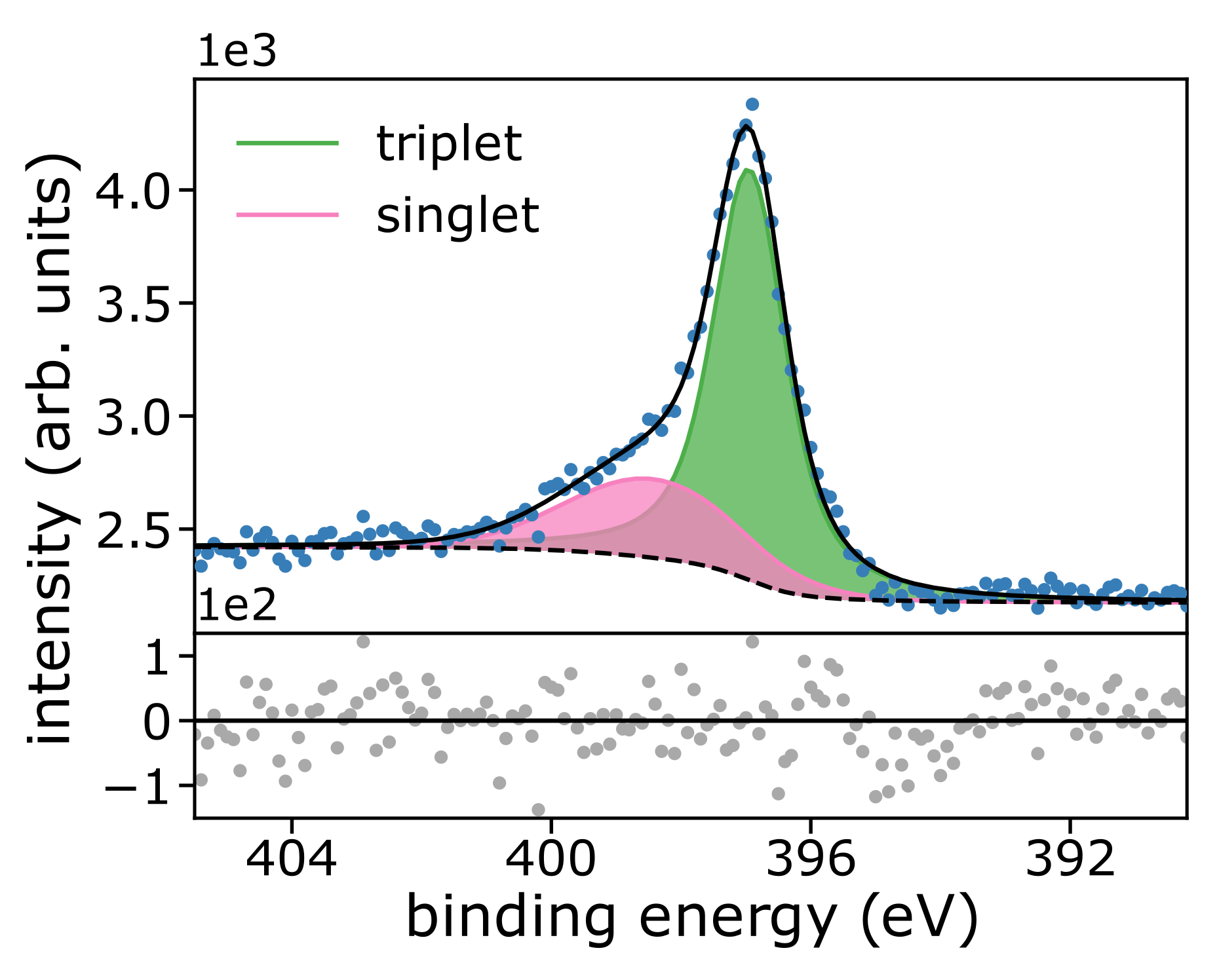}
	\caption{\label{fig:XPS} N 1s X-ray photoelectron spectroscopy (XPS) of NiO\textsubscript{1-x}N\textsubscript{x} with incoming photon energy of 1486.7 eV (Al K-alpha). The triplet to singlet ratio is $2.8\pm0.3$. The peaks and backgrounds are fitted simultaneously using the Shirley-Vegh-Salvi-Castle (SVSC) background (peak Shirley) \cite{Herrera-Gomez2014}.
	}
\end{figure}

The N 1s XPS measurement shown in Fig. \ref{fig:XPS} allows us to confirm the nature of the hole in N-substituted NiO. It shows a double peak contribution similar to other oxynitrides \cite{Elfimov2007a}. In NiO\textsubscript{0.94}N\textsubscript{0.06}, we find a peak separation of 1.5$\pm$0.1 eV and an intensity ratio of 2.8$\pm$0.3, consistent with the expected triplet to singlet 3:1 ratio. This points to a coupling between the N 2p electrons and the core hole (a single peak would be observed otherwise), confirming the N 2p nature of the introduced holes. All N concentrations in this manuscript are calculated by comparing the intensity ratios and atomic sensitivity factors of N 1s and Ni 2p \cite{Moulder1995a} (see Supplemental Material \cite{SM_PRL} for additional XPS data).

The combined spectroscopy results in this Letter complete our understanding of NiO\textsubscript{1-x}N\textsubscript{x} and corroborate the properties predicted by DFT. The calculations reveal that N substitution significantly perturbs the local magnetic environment, modifying the magnetic moment at neighboring Ni cation sites and giving rise to the formation of Ni-N-Ni magnetic centers. Experimentally, the suppressed dichroism at the Ni L\textsubscript{2}-edge supports this lowest energy configuration. Comparing the experimental and calculated N K-edge XAS confirms the in-gap impurity states of N 2p character predicted by DFT. The additional states above the valence band yield an emerging band in ARPES, with intensity scaling linearly with N content. The hole-like nature of the N-derived states is further corroborated by N 1s core-level XPS.

In-gap transitions within Ni-N-Ni centers introduces the possibility of leveraging these defect states for quantum device applications. Thus, one must consider all spin configurations involving both Ni high-spin 3d electrons ($S_{Ni-Ni} = 2,1,0$) and the N 2p hole ($S_{N} = 1/2$). This yields the following: one $S_{tot} = 5/2$, three $S_{tot} = 3/2$, and two $S_{tot} = 1/2$. Transitions that conserve total spin ($S_{tot} = 3/2 \rightarrow S_{tot} = 3/2$) are allowed, with the $S_{tot} = 3/2$ initial state dictated by the favored Ni-Ni FM double exchange and Ni-N AFM exchange ground state. The energy splitting of these states is on the order of the exchange energy ($\sim 250$ meV), well within the energy scale of the bandgap of NiO. Furthermore, spin-orbit coupling is expected to be significant, given the spin-orbit constant of $\sim 40$ meV of Ni 3d\textsuperscript{8} \cite{Radwanski2000}, enabling additional transitions ($S_{tot} = 3/2 \rightarrow S_{tot} = 5/2, 1/2$).

In conclusion, the discovery of novel magnetic behavior resulting from individual N substitution in NiO reveals a new and largely unexplored avenue for controlling local magnetism in strongly correlated oxides via anion substitution. The formation of Ni–NNi centers introduces localized spin structures with potential quantum functionality, opening opportunities for defect engineering in magnetic materials. To assess their technological viability, particularly for quantum sensing, further studies are needed on the growth and behavior of highly diluted films (N concentration $<1$\%). A critical question remains whether these defect centers are magnetically decoupled from the AFM lattice and host optically or magnetically addressable in-gap transitions. Temperature is an important variable, as magnetic disorder above 0 K induces a net magnetic field from the surrounding AFM lattice. This may be advantageous, since thermally excited magnons cause local spin fluctuations and could mediate a short-range coupling between two nearby Ni-N-Ni centers. In NV centers, such magnon-induced coupling requires a heterostructure of diamond and a magnetic material \cite{Jiang2017}, but could be intrinsic within NiO.

\begin{acknowledgments}

\vspace{0.4cm}

\textit{Acknowledgement}---This research was undertaken thanks in part to funding from the Max Planck-UBC-UTokyo Center for Quantum Materials and the Canada First Research Excellence Fund, Quantum Materials and Future Technologies Program. The work at the University of British Columbia was also supported by the Canada Foundation for Innovation (CFI) and the British Columbia Knowledge Development Fund (BCKDF). Part of the research described in this Letter was performed at the Canadian Light Source, a national research facility of the University of Saskatchewan, which is supported by the Canada Foundation for Innovation (CFI), the Natural Sciences and Engineering Research Council (NSERC), the Canadian Institutes of Health Research (CIHR), the Government of Saskatchewan, and the University of Saskatchewan. This research used resources of the Advanced Light Source, which is a DOE Office of Science User Facility under Contract No. DE-AC02-05CH11231.

\vspace{0.4cm}

\textit{Data availability}---The data that support the findings of this article are not publicly available upon publication because it is not technically feasible and/or the cost of preparing, depositing, and hosting the data would be prohibitive within the terms of this research project. The data are available from the authors upon reasonable request.

\end{acknowledgments}



%

\end{document}


\title{Supplementary Material for: Novel Magnetic Ni-N-Ni Centers in N-substituted NiO}

\author{Simon Godin}
\email{Corresponding author: sgodin@phas.ubc.ca}
\affiliation{Department of Physics and Astronomy, University of British Columbia, Vancouver, British Columbia V6T 1Z4, Canada}
\affiliation{Quantum Matter Institute, University of British Columbia, Vancouver, British Columbia V6T 1Z4, Canada}

\author{Ilya S. Elfimov}
\affiliation{Department of Physics and Astronomy, University of British Columbia, Vancouver, British Columbia V6T 1Z4, Canada}
\affiliation{Quantum Matter Institute, University of British Columbia, Vancouver, British Columbia V6T 1Z4, Canada}

\author{Fengmiao Li}
\affiliation{Department of Physics and Astronomy, University of British Columbia, Vancouver, British Columbia V6T 1Z4, Canada}
\affiliation{Quantum Matter Institute, University of British Columbia, Vancouver, British Columbia V6T 1Z4, Canada}

\author{Bruce A. Davidson}
\affiliation{Department of Physics and Astronomy, University of British Columbia, Vancouver, British Columbia V6T 1Z4, Canada}
\affiliation{Quantum Matter Institute, University of British Columbia, Vancouver, British Columbia V6T 1Z4, Canada}

\author{Ronny Sutarto}
\affiliation{Canadian Light Source, Saskatoon, Saskatchewan S7N 2V3, Canada}

\author{Jonathan D. Denlinger}
\affiliation{Advanced Light Source, Lawrence Berkeley National Laboratory, Berkeley, CA 94720, USA}

 \author{Liu Hao Tjeng}
\affiliation{Max Planck Institute for Chemical Physics of Solids, Nöthnitzer Straße 40, 01187 Dresden, Germany }

\author{George A. Sawatzky}
\affiliation{Department of Physics and Astronomy, University of British Columbia, Vancouver, British Columbia V6T 1Z4, Canada}
\affiliation{Quantum Matter Institute, University of British Columbia, Vancouver, British Columbia V6T 1Z4, Canada}

\author{Ke Zou}
\affiliation{Department of Physics and Astronomy, University of British Columbia, Vancouver, British Columbia V6T 1Z4, Canada}
\affiliation{Quantum Matter Institute, University of British Columbia, Vancouver, British Columbia V6T 1Z4, Canada}

\date{\today}

\maketitle

\section{Thin Film Fabrication \label{sec:structure}}

\begin{figure}[th!]
	\includegraphics[width = \columnwidth]{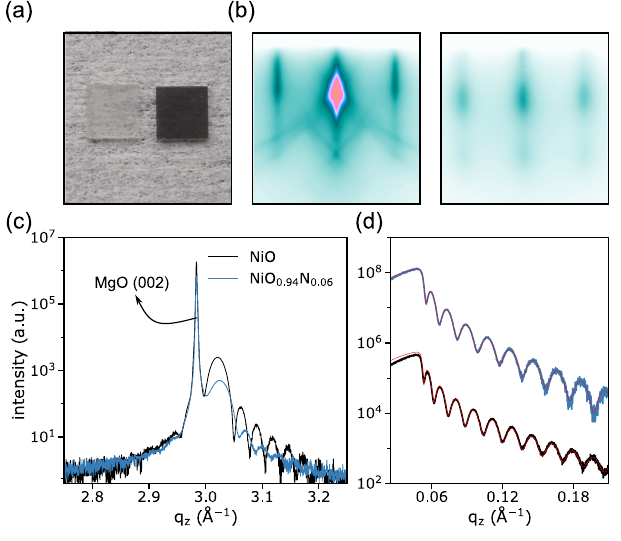}
	\caption{\label{fig:structure} Structure and crystal quality of NiO and NiO\textsubscript{0.94}N\textsubscript{0.06}. (a) Picture of NiO and NiO\textsubscript{0.94}N\textsubscript{0.06} samples. (b) RHEED images of NiO (001) and NiO\textsubscript{0.94}N\textsubscript{0.06} (001) in the [100] direction. (c) XRD spectra of NiO and NiO\textsubscript{0.94}N\textsubscript{0.06}, showing thickness fringes for both films. Note the decreased lattice parameter upon N substitution. The scans are shifted to match the baseline of both backgrounds.  (d) XRR spectra of NiO and NiO\textsubscript{0.94}N\textsubscript{0.06}. The simulations in red show a thickness of 347$\pm$1 and 297.0$\pm$0.7 \SI{}{\angstrom}, respectively.
    }
\end{figure}

In this Section, we describe the growth and characterization of NiO and N-substituted NiO. N-substituted NiO films have previously been synthesized using deposition techniques, such as radio-frequency magnetron sputtering in N\textsubscript{2} environment\cite{Keraudy2017a,Lin2013a}, or spray pyrolysis\cite{Sriram2016b}. These techniques generate polycrystalline or amorphous films. However, single-crystalline synthesis of substituted films is desirable for complete characterization and necessary to limit undesired defects and disorder. We instead use nitrogen oxide (NO) gas-assisted molecular beam epitaxy (MBE), which allows us to have fine control over the substitution concentration and high crystallinity. As in other compounds\cite{Wicks2012b,Voogt2001a}, the inclusion of N is achieved by lowering the NO gas pressure and the substrate temperature during growth. This would normally lead to O deficiency in the system, but here allows for the substitution of the O sites by N instead.

The films presented in the main text are grown epitaxially on MgO (001). We use MgO as a substrate due to its rock-salt structure and the $<$ 1\% lattice mismatch from NiO \cite{Godin2022}. The substrates are cleaned with ethanol and methanol and annealed in ultra-high vacuum (UHV) at \SI{800}{\celsius}. In our apparatus, the NO gas is carried from a leak valve by a feedthrough directed at the sample's surface to raise the effective flux. This method allows good control over the N substitution level by varying the growth temperature and NO gas pressure. The films presented here are grown at \SI{150}{\celsius} and NO pressure ranging from \SI{1e-6}{} to \SI{4e-7}{Torr}. This results in N atom concentrations ranging from 6 to 12\%. The unsubstituted NiO control sample is grown at \SI{500}{\celsius} in molecular O pressure of \SI{5e-6}{Torr}. We compare exemplary unsubstituted NiO and 6\% substituted NiO\textsubscript{1-x}N\textsubscript{x} films in Fig. \ref{fig:structure}. We see from Fig. \ref{fig:structure}(a) that the films become darker with substitution. This effect, along with the increased conductivity of the samples, provides a useful way to quickly confirm the presence of N.

Reflection high-energy electron diffraction (RHEED) measurements confirm the same rock-salt structure of NiO\textsubscript{1-x}N\textsubscript{x} as NiO [Fig. \ref{fig:structure}(b)]. The substituted films carry different features in RHEED, i.e. modulated streaks. This indicates higher surface roughness, although the strong streak intensity implies the single orientation and good quality of the films at the surface. High-resolution X-ray diffraction scans (HRXRD) show high crystallinity in the bulk of both samples [Fig. \ref{fig:structure}(c)], indicated by the presence of thickness fringes. Both NiO\textsubscript{1-x}N\textsubscript{x} and NiO have tensile strain on the MgO substrate, leading to a smaller lattice parameter in the c direction and a tetragonal unit cell, due to the biaxial strain. N substitution also slightly reduces the c lattice parameter. The XRR scans in Fig. \ref{fig:structure}(d) show clear oscillations for both films. The thickness of the samples, ranging from 30 to 35 nm, is determined by simulation of the XRR spectra using the GenX software\cite{Bjorck2007}.

\begin{figure*}[]
	\includegraphics[width = \textwidth]{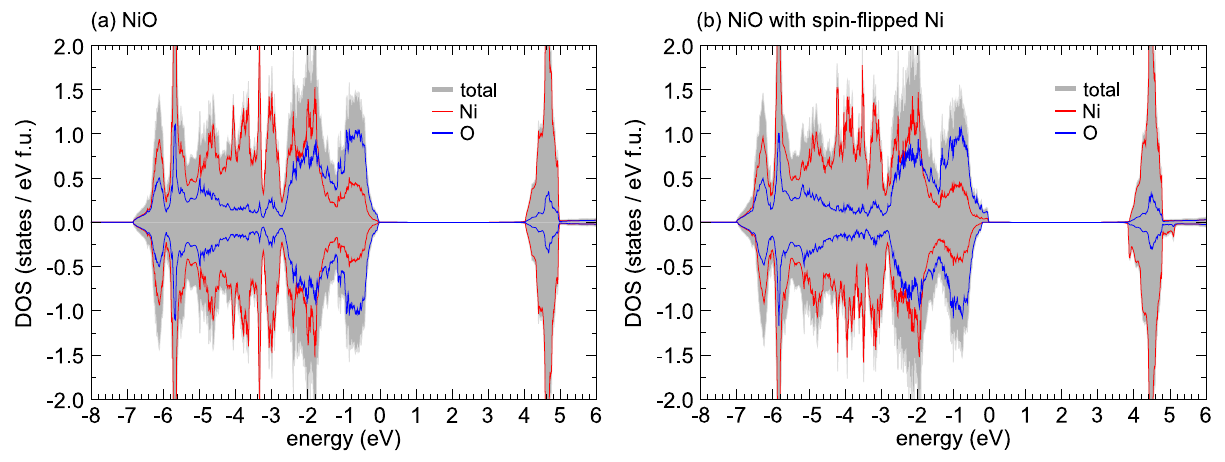}
	\caption{\label{fig:DFT_NiO_no_spin-flip_NiO} Additional DFT calculations for different configurations in an unsubstituted NiO unit cell with 32 anion and cation sites. PDoS of (a) NiO and (b) NiO with one spin-flipped Ni. The cost to flip the magnetic moment of the Ni is 117 meV. The conductivity gap is reduced to 3.44 eV from 3.63 eV between both configurations. The zero of energy is at the Fermi energy.
	}
\end{figure*}

\begin{figure*}[]
	\includegraphics[width = \textwidth]{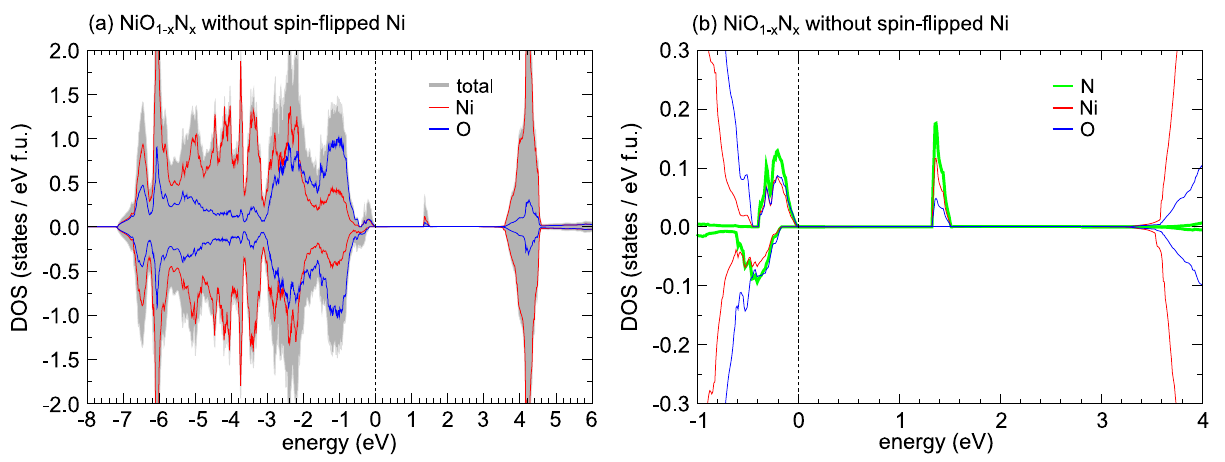}
	\caption{\label{fig:DFT_NiO_no_spin-flip_NiON} Additional DFT calculations for different configurations in an N-substituted NiO unit cell with 32 anion and cation sites. (a) PDoS of NiO\textsubscript{1-x}N\textsubscript{x} with one O substituted for N, but without the spin-flipped Ni configuration. This solution raises the energy of the system by 204 meV. (b) Zoom of (a) around the bandgap, showing the N states. The zero of energy is at the Fermi energy. The N states are not shown in (b).
	}
\end{figure*}

\clearpage

\section{DFT Calculations and Additional Configurations}

\begin{figure}[b]
	\includegraphics[width = 0.8\columnwidth]{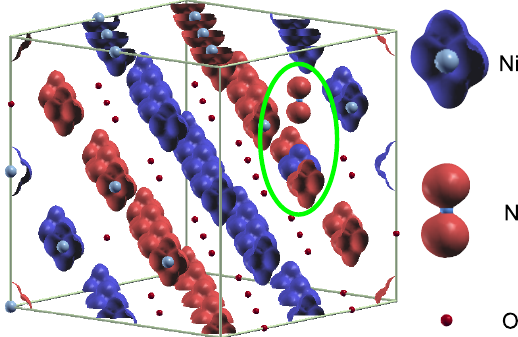}
	\caption{\label{fig:spin_density} (c) Spin density showing the orientation of N spin density with respect to Ni site whose magnetic moment is changed with respect to bulk NiO AFM structure. Spin-up and spin-down states are indicated in blue and red, respectively.
    }
\end{figure}

The spin-polarized DFT calculations for spin-flipped NiO\textsubscript{1-x}N\textsubscript{x} in the main text are performed with the Quantum Espresso code\cite{QE-2009,QE-2017}, using the projector augmented wave (PAW) method\cite{PAW-1994}. The Perdew-Burke-Ernzerhof (PBE) functional\cite{PBE-1996} is used for the exchange-correlation energy. The kinetic energy cutoff for wavefunctions is 1088.5 eV. The Brillouin zone of 2x2x2 supercell is sampled using 8x8x8 k-grid. The Dudarev et al. formulation of DFT+U method\cite{DFTU-1991,DFTU-1998} is used throughout with Ni 3d, O 2p and N 2p electrons’ on-site effective interactions U = 6 eV, 4 eV, and 3 eV, respectively. The calculated bulk NiO lattice constant of 4.24 \SI{}{\angstrom} and the band gap of 3.63 eV are in good agreement with experimental observations. N XAS spectra are calculated using the theoretical approach after Gougoussis et al.\cite{kedges1,kedges2}.

Additional PDoS from DFT calculations are shown in Fig. \ref{fig:DFT_NiO_no_spin-flip_NiO} and Fig. \ref{fig:DFT_NiO_no_spin-flip_NiON}. The energy cost of flipping a Ni spin in unsubstituted NiO is 117 meV [Fig. \ref{fig:DFT_NiO_no_spin-flip_NiO}(a) and Fig. \ref{fig:DFT_NiO_no_spin-flip_NiO}(b)]. On the other hand, the arrangement with no Ni spin-flip in NiO\textsubscript{1-x}N\textsubscript{x} [Fig. \ref{fig:DFT_NiO_no_spin-flip_NiON}(a) and Fig. \ref{fig:DFT_NiO_no_spin-flip_NiON}(b)] raises the ground state energy by 204 meV compared to the one in the main text. While the N atom in the Ni spin-flipped configuration stays in the center of the Ni octahedral, this is not the case for this high-energy state. The N atom moves along the (111) plane, resulting in Ni-N interatomic distances of 4.361 \SI{}{\angstrom} in the z direction and 4.207 \SI{}{\angstrom} in the x and y direction. Figure \ref{fig:spin_density} shows the spin-density map of this system. It allows for easy visualization of the N hole and spin-flipped Ni pair, with the N orientated towards the Ni. This changes the AFM Ni-O-Ni into Ni-N-Ni with both Ni spins in the same direction, opposite to the spin of the N hole.

\begin{figure}[t]
	\includegraphics[width = \columnwidth]{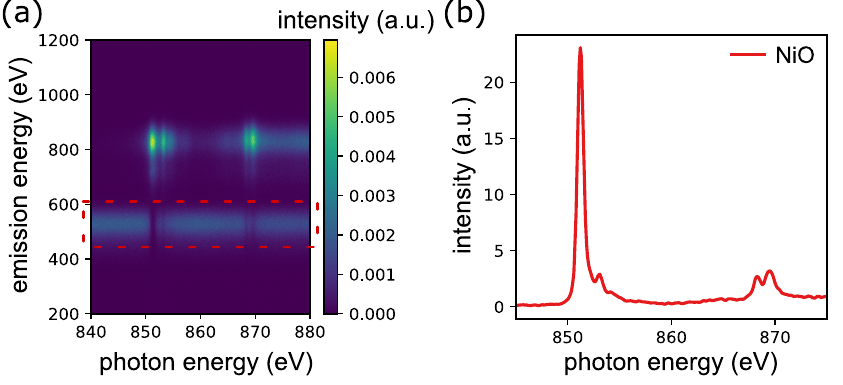}
	\caption{\label{fig:IPFY} XAS IPFY spectrum of the Ni L\textsubscript{3,2}-edges in NiO. (a) Photon vs. emission energy map. The Ni L\textsubscript{3,2}-edges (850-870 eV) show self-absorption effects. The O 1s contribution (530-580 eV) in the red dashed rectangle are summed over and inverted to obtain the IPFY data (b).
    }
\end{figure}

\begin{figure}[t]
	\includegraphics[width = 0.8\columnwidth]{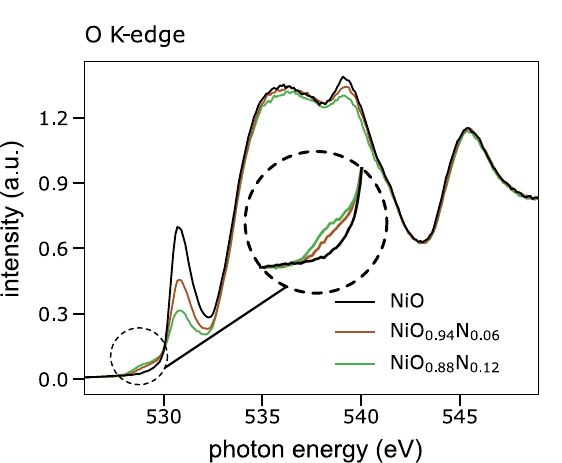}
	\caption{\label{fig:XAS} XAS of NiO\textsubscript{1-x}N\textsubscript{x} at the O K-edge in TFY mode as a function of substitution, which suppresses the pre-edge and introduces a shoulder (blown-up region).
    }
\end{figure}

\section{XAS and Calculated N K-edge}

\begin{figure*}[t]
	\includegraphics[width = 0.8 \textwidth]{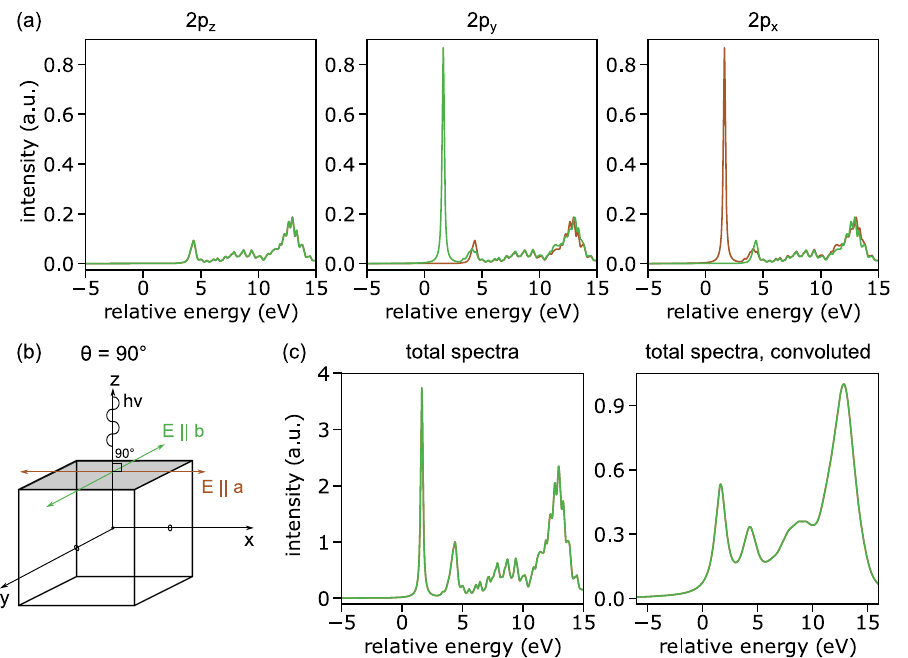}
	\caption{\label{fig:appendixB_all} Calculated XAS N K-edge from the DFT results in the main text. (a) Calculated N K-edge, resolved for p\textsubscript{z}, p\textsubscript{y} and p\textsubscript{x}. (b) Geometry of the polarization of the incoming light at normal incidence. The shaded area shows the surface of the sample. (c) Total spectra, with and without a 1 eV Lorentzian broadening.
	}
\end{figure*}

The XAS measurements in the main text are conducted under linear polarized light. At normal incidence ($\theta$ = \SI{90}{\degree}), the polarization vectors are along E $||$ a and E $||$ b. Grazing incidence ($\theta$ = \SI{30}{\degree}) leads to (2/3 E $||$ c + 1/3 E $||$ a) and E $||$ b. This geometry is shown in Fig. \ref{fig:appendixB_all}(b). We use three modes of detection, namely total fluorescence and electron yield (TFY and TEY), as well as inverse partial fluorescence yield (IPFY)\cite{Achkar2011a}. Each mode comes with its own set of advantages and is used in this paper.

The TEY mode offers a shallower penetration depth and a spectrum closer to the actual absorption coefficient compared to the TFY mode. This is due to the self-absorption effects that populate the fluorescence yield, originating from additional absorption at the edge’s resonant energies. However, the TEY mode typically suffers from significant charging effects due to the insulating nature of NiO. This makes it particularly difficult to compare the substituted films to unsubstituted NiO since they have a large difference in resistivity. Lastly, the IPFY mode resolves outgoing photon energy and uses the contribution of a non-resonant transition at lower detected energy to extract the spectra without including charging or self-absorption effects. With this in mind, we choose to analyze the O K-edge using TFY, since we compared NiO\textsubscript{1-x}N\textsubscript{x} and NiO. The use of both TEY and TFY for the N K-edge is ideal due to the limited difference in charging effects between different substitution levels, as well as allowing the comparison of different probing depths. For the Ni L\textsubscript{3,2}-edges, the high intensity of the O 1s states at lower emission energy allows for the use of IPFY, as shown in Fig. \ref{fig:IPFY}. This leads to spectra that are the closest to the absorption coefficient.

The process to obtain the IPFY XAS spectrum of the Ni L\textsubscript{3,2}-edges in NiO is shown in Fig. \ref{fig:IPFY}. The photon vs. emission energy map in Fig. \ref{fig:IPFY}(a) contains both the Ni 2p and O 1s states, around 850 and 550 eV, respectively. The intensity near Ni 2p includes self-absorption effects, while the dip in intensity near O 1s reflects the absorption coefficient without such effects. The resulting Ni L\textsubscript{3,2}-edges absorption is seen in Fig. \ref{fig:IPFY}(b), by taking the inverse of the O 1s states in the red dashed rectangle in (a).

The TFY spectra at the O K-edge for 0, 6, and 12\% substituted films are shown in Fig. \ref{fig:XAS}. A pre-edge is present at 530.5 eV, due to hybridization between Ni 3d and O 2p states. The reduction in the pre-edge intensity as a function of substitution shows a larger ratio than 6 and 12\%. This can be understood by the fact that each N has 6 Ni nearest-neighbors, affecting their hybridization with O significantly. A new peak also emerges at 529 eV, emphasized by the dashed circle, originating from the N 2p states introduced in the bandgap. The intensity is higher with increasing N doping levels.

Figure \ref{fig:appendixB_all}(a) shows the calculated XAS spectra for the 1s to 2p transitions of the N K-edge, resolved for 2p\textsubscript{z}, 2p\textsubscript{y} and 2p\textsubscript{x}. The geometry of the experimental and calculated N K-edge are identical. Note that at normal incidence, the 2p\textsubscript{y} and 2p\textsubscript{x} spectra are identical, but with reversed polarization, as expected from a cubic system. Figure \ref{fig:appendixB_all}(c) shows the resulting total spectra. A 1 eV Lorentzian broadening leads to good agreement with the experimental data.

\section{ARPES and XPS Analysis}

\begin{figure}[t]
	\includegraphics[width = \columnwidth]{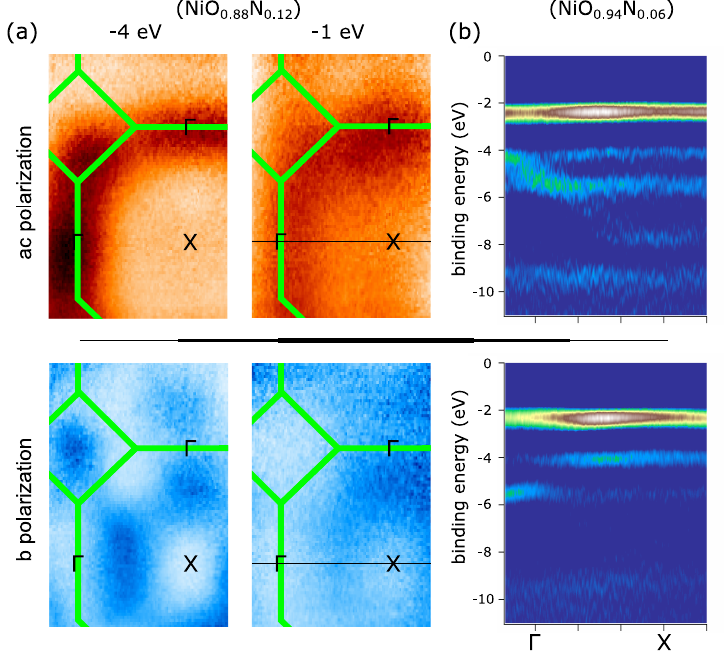}
	\caption{\label{fig:ARPES} ARPES on NiO\textsubscript{1-x}N\textsubscript{x} with incoming photon energy of 85 eV and grazing incidence ($\theta$ = \SI{25}{\degree}). (a) Constant-energy maps at binding energies of -4 and -1 eV with ac and b polarizations. The 12\% substituted sample is shown to resolve the N states at -1 eV. (b) Momentum Distribution Curve (MDC) 2\textsuperscript{nd} derivative of $\Gamma$-$X$ cuts in NiO\textsubscript{0.94}N\textsubscript{0.06}.
    }
\end{figure}

\begin{figure}[t]
	\includegraphics[width = \columnwidth]{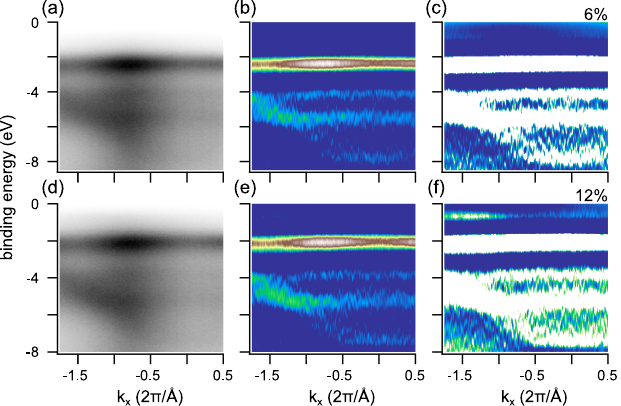}
	\caption{\label{fig:2ndDerivative} Comparison of the $\Gamma$-$X$ cuts in the ARPES spectra of 6 and 12\% substituted films for raw data (a, d) and 2\textsuperscript{nd} derivative (b, e). Taking the logarithm scale in intensity of the 2\textsuperscript{nd} derivative (c, f) emphasizes the N 2p band, located around -0.75 eV.
    }
\end{figure}

We present here additional ARPES and XPS results for NiO\textsubscript{1-x}N\textsubscript{x}. Fig. \ref{fig:ARPES}(a) shows constant-energy maps of NiO\textsubscript{0.88}N\textsubscript{0.12} at a binding energy of -4 eV (crossing of the hybridized O 2p and Ni 3d bands) and -1 eV (N 2p states in the valence band). These scans are taken \textit{ex situ}, after a simple degassing procedure at \SI{150}{\celsius}. Our ability to observe clear constant-energy maps confirms the high crystallinity and resilience to air exposure of the substituted samples. The incoming photon energy is 85 eV and the scans are taken at grazing incidence ($\theta$ = \SI{25}{\degree}) in a similar geometry than XAS (E $||$ a and E $||$ b at normal incidence).

Figure \ref{fig:ARPES}(b) shows the cuts along $\Gamma$ to $X$ of NiO\textsubscript{0.94}N\textsubscript{0.06} under linear polarized light. The high-intensity Ni 3d band is seen at -2.2 eV, while the hybridized states of O 2p and Ni 3d character spread between -4 to -8 eV. A clear polarization dependence of the lowest band is observed through this whole region, indicative of out-of-plane symmetry. A core-level band is also present at -9.5 eV. Figure \ref{fig:2ndDerivative} compares the raw data [Fig. \ref{fig:2ndDerivative}(a) and \ref{fig:2ndDerivative}(d)] and the 2\textsuperscript{nd} derivative [Fig. \ref{fig:2ndDerivative}(b) and Fig. \ref{fig:2ndDerivative}(e)] in the case of 6\% and 12\% substitution. The 2\textsuperscript{nd} derivative is used to accurately determine the maximum intensity of the bands and reduce the background of the detector. Figure \ref{fig:2ndDerivative}(c) and Fig. \ref{fig:2ndDerivative}(f) show the logarithm scale of the intensity of the 2\textsuperscript{nd} derivative. This allows us to resolve the N 2p band for 12\% substitution, located around -0.75 eV. The band is faint in the case of 6\% substitution, but still visible around the $\Gamma$ and $X$ points.

\begin{figure}[t]
	\includegraphics[width = \columnwidth]{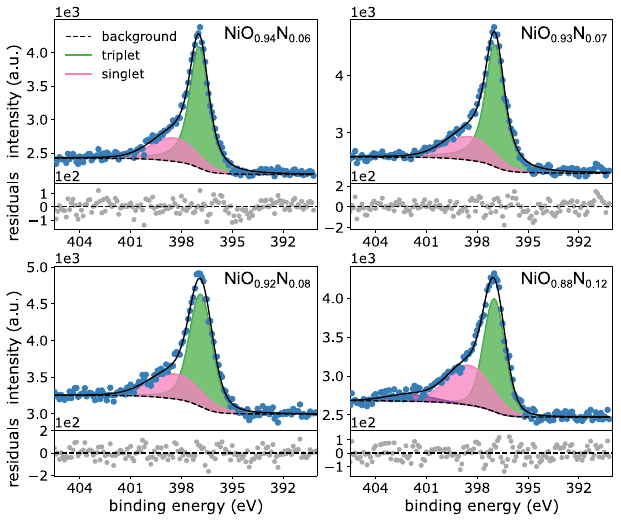}
	\caption{\label{fig:XPS_N1s} XPS spectra of N 1s for 6\% 7\%, 8\% and 12\% substitution. The triplet to singlet ratios are $2.9\pm0.3$, $2.9\pm0.3$, $2.3\pm0.3$ and $1.5\pm0.4$, with a $\chi^2_{\text{r}}$ of 2.22, 1.48, 1.9 and 1.14 respectively. The peaks and backgrounds are fitted simultaneously using the SVSC background (peak Shirley)\cite{Herrera-Gomez2014}
    }
\end{figure}

Results for the XPS of the N 1s samples are shown in Fig. \ref{fig:XPS_N1s}. Note that the 3:1 ratio is reduced at higher substitution. A shake-off peak also becomes visible at 12\%. 


%